\documentclass[10pt]{iopart}
\usepackage{units}
\usepackage{graphicx}
\usepackage{epstopdf}
\usepackage{color}

\usepackage{hyperref}

\begin{document}

\title[Superconducting Sn films above and below percolation]{Microwave study of superconducting Sn films above and below percolation}

\author{Manfred H. Beutel, Nikolaj G. Ebensperger, Markus Thiemann, Gabriele Untereiner, Vincent Fritz, Mojtaba Javaheri, Jonathan N\"agele, Roland R\"osslhuber, Martin Dressel, and Marc Scheffler}

\address{1. Physikalisches Institut, Universit\"at Stuttgart, Pfaffenwaldring 57, D-70569 Stuttgart, Germany}
\ead{scheffl@pi1.physik.uni-stuttgart.de}
\vspace{10pt}
\begin{indented}
\item[] \today
\end{indented}

\begin{abstract}
The electronic properties of superconducting Sn films ($T_c \approx \unit[3.8]{K}$) change significantly when reducing the film thickness down to a few $\unit{nm}$, in particular close to the percolation threshold. The low-energy electrodynamics of such Sn samples can be probed via microwave spectroscopy, e.g.\ with superconducting stripline resonators. 

Here we study Sn thin films, deposited via thermal evaporation -ranging in thickness between $\unit[38]{nm}$ and $\unit[842]{nm}$- which encompasses the percolation transition.
We use superconducting Pb stripline resonators to probe the microwave response of these Sn films in a frequency range between $\unit[4]{GHz}$ and $\unit[20]{GHz}$ at temperatures from $\unit[7.2]{K}$ down to $\unit[1.5]{K}$. The measured quality factor of the resonators decreases with rising temperature due to enhanced losses. As a function of the sample thickness we observe three regimes with significantly different properties: samples below percolation, i.e.\ ensembles of disconnected superconducting islands, exhibit dielectric properties with negligible losses, demonstrating that macroscopic current paths are required for appreciable dynamical conductivity of Sn at GHz frequencies. 
Thick Sn films, as the other limit, lead to low-loss resonances both above and below $T_c$ of Sn, as expected for bulk conductors. But in an intermediate thickness regime, just above percolation and with labyrinth-like morphology of the Sn, we observe a quite different behavior: the superconducting state has a microwave response similar to the thicker, completely covering films with low microwave losses; but the metallic state of these Sn films is so lossy that resonator operation is suppressed completely.

\end{abstract}
\ioptwocol

\section{Introduction}
Microwave spectroscopy offers experimental access to the low-energy electrodynamics of solids and can reveal important information about a wide variety of material classes \cite{dressel_buch,Schwartz2000,Lunkenheimer2000,Bovtun2001,scheffler_slow_drude_relaxation_2005,Krupka2006,dressel_scheffler_ann_phys_2006}. 
Particularly, superconductors with their superconducting gap as characteristic energy scale in the range of $\unit{\mu eV}$ to $\unit{meV}$ are intensively studied with microwaves \cite{Bonn1992,Kokales2000,Steinberg2008,Hashimoto2009,Pompeo2010}. 
The numerous experimental challenges involved in such studies (for example sensitivity, frequency range, temperature range, sample dimensions) cannot be simultaneously overcome by a single technique, and therefore
several experimental approaches have been developed \cite{Klein1993,Booth1994,bolometric_approach_Turner_Broun_2004,Scheffler2005a,Huttema2006,scheffler_broadband_planar/stripline_resonators_corbino_2012,scheffler_corbino_stripline_2015}. One such method consists of resonant measurements using stripline geometry, where the sample acts as perturbation to the standing microwave pattern forming in the resonator \cite{dilorio_stripline_1988,Oates1990,Revenaz1994,Belk1996,Oates2010,Scheffler2012,hafner_2014,markus_all_Nb_and_Ta_loaded_resonator_2014}. This is a sensitive technique which gives access to complex physical quantities like surface impedance and conductivity \cite{dilorio_stripline_1988,Oates1990,hafner_2014}. Stripline resonators can be easily operated at different discrete frequencies and therefore the spectral behavior of the material under study is accessible. Furthermore, they are compatible with typical cryogenic apparatus, which gives access to the low temperature properties, down to the mK range, of bulk as well as thin-film superconductors \cite{scheffler_corbino_stripline_2015,dilorio_stripline_1988,hafner_2014, HafnerJPS_2014, Parkkinen2015}.
 
Here we are interested in the electrodynamics of superconducting thin films with varying thickness, a research field that has recently attracted a lot of attention in the context of the superconductor-to-insulator transition (SIT) observed in strongly, homogeneously disordered films of materials such as InO$_x$, TiN, or NbN \cite{pfuner_opt_prop_2009,liu_corbino_2011,driessen_2012,Pracht2012,matsunaga_2013,sherman_pracht_2015}.
In the present work, we also touch upon a transition between a superconducting state and a (macroscopically) insulating state as a function of thickness, but this transition is of percolative nature and thus follows a completely different mechanism \cite{Kirkpatrick1973,Deutscher1980}. Superconductors with properties that depend on percolation comprise various thin-film, polycrystalline, composite and cluster-assembled materials \cite{Deutscher1980,Golosovsky1992,Eisterer2003,Yamamoto2007,Cuppens2010}.
While the percolation transition in the growth of metallic thin films depends on the evolution of their microstructure and can be identified with DC transport measurements, the electrodynamic response of such films close to percolation is of great interest throughout a wide spectral range up to optical frequencies as well \cite{Clerc1990,Dyre2000,Hövel2010}.

The main motivation of the present work, however, is not the percolation transition itself or possible direct signatures thereof in the dynamical response, but instead the role that the thickness of superconducting films plays when such films are part of certain microwave elements, namely superconducting stripline resonators.
Considering conductive thin films in general, many of the electronic properties change significantly with decreasing thickness. In the case of thermally evaporated Sn, this is accompanied by structural transitions, e.g. island structure, formation of aggregates, labyrinth structure, and continuous films for increasing film thickness \cite{andrew_critfield_1948,lock_1951,blumberg1962,hall_1965,increaseTc_1970,dolan1974,eichele1981}. 
It is expected that structural transitions of Sn will lead to changes in electronic scattering, which in turn will influence the electrodynamic properties of the sample. Thus conductivity and resistivity, both for DC and microwaves, will strongly depend on thickness. 

In this system of superconducting Sn films at GHz frequencies, there are three different relevant length scales, namely superconducting penetration depth $\lambda$, skin depth $\delta$ and film thickness $d$. As a function of temperature $T$, both $\lambda$ and $\delta$ will change while $d$ remains constant. Considering additionally the critical film thickness $d_c$ of the percolation threshold, one has a complex system in which the behavior of electrodynamic quantities like resistivity $\rho(\omega,T)$ and therefore the behavior of the stripline resonator will depend on the relative size of these length scales in an intricate fashion. Slightly above percolation high losses are expected due to increased disorder of the film, which should decrease by increasing film thickness. In the state below percolation with disconnected islands an infinite DC resistivity $\rho_\mathrm{DC} \rightarrow \infty$ is expected. It could be possible to observe non-zero dynamical conductivity when the island size is greater than the oscillatory amplitude of the conduction electrons which is governed by the microwave frequency. Therefore a frequency-dependent behavior, resulting in enhanced conductivity with increasing frequency, is conceivable.

In this article we present a study of thermally evaporated Sn films above and below percolation using superconducting Pb stripline resonators. With this contactless measurement method the microwave response of different samples ranging in thickness between  $\unit[37]{nm}$ and $\unit[842]{nm}$ are measured. Furthermore, the samples are characterized by imaging the surface topography using atomic force microscopy (AFM), and in terms of conductivity DC measurements are performed.

\begin{figure*}[tbph]
\centering
\includegraphics[width=0.32\linewidth]{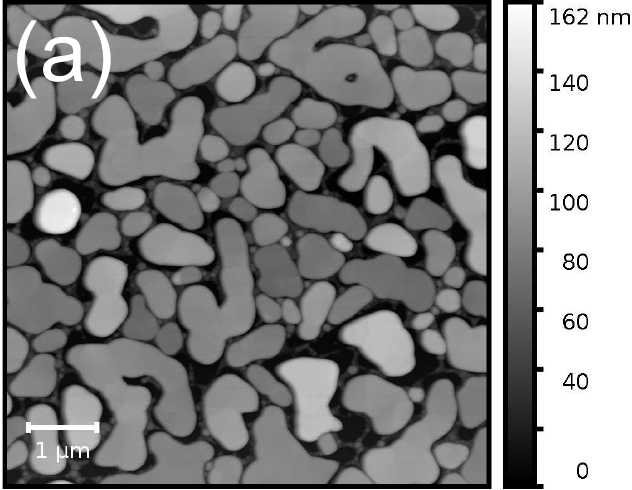}\hfill
\includegraphics[width=0.32\linewidth]{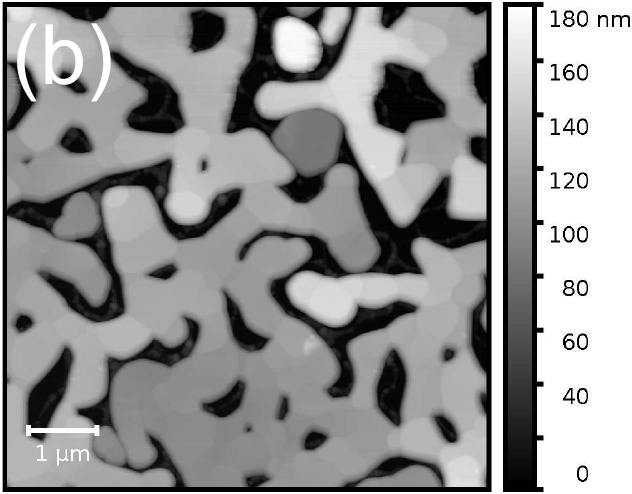}\hfill
\includegraphics[width=0.32\linewidth]{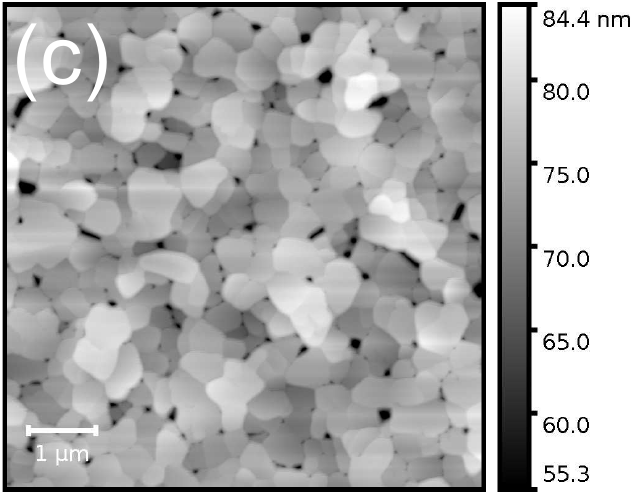}
\caption{AFM images of thickness $d=\unit[68]{nm}$ (a), $\unit[93]{nm}$ (b) and $\unit[348]{nm}$ (c) showing the surface topography of Sn. It is seen that Sn starts to grow as isolated, disconnected islands (a) and by increasing the film thickness the condensate forms a labyrinth structure (b). For the sample $d=\unit[348]{nm}$ the islands overlap and the sample forms a continuously connected surface.}
\label{fig:afm}
\end{figure*}

\section{Samples}
\begin{figure}[tbph]
\centering
\includegraphics[width=\linewidth]{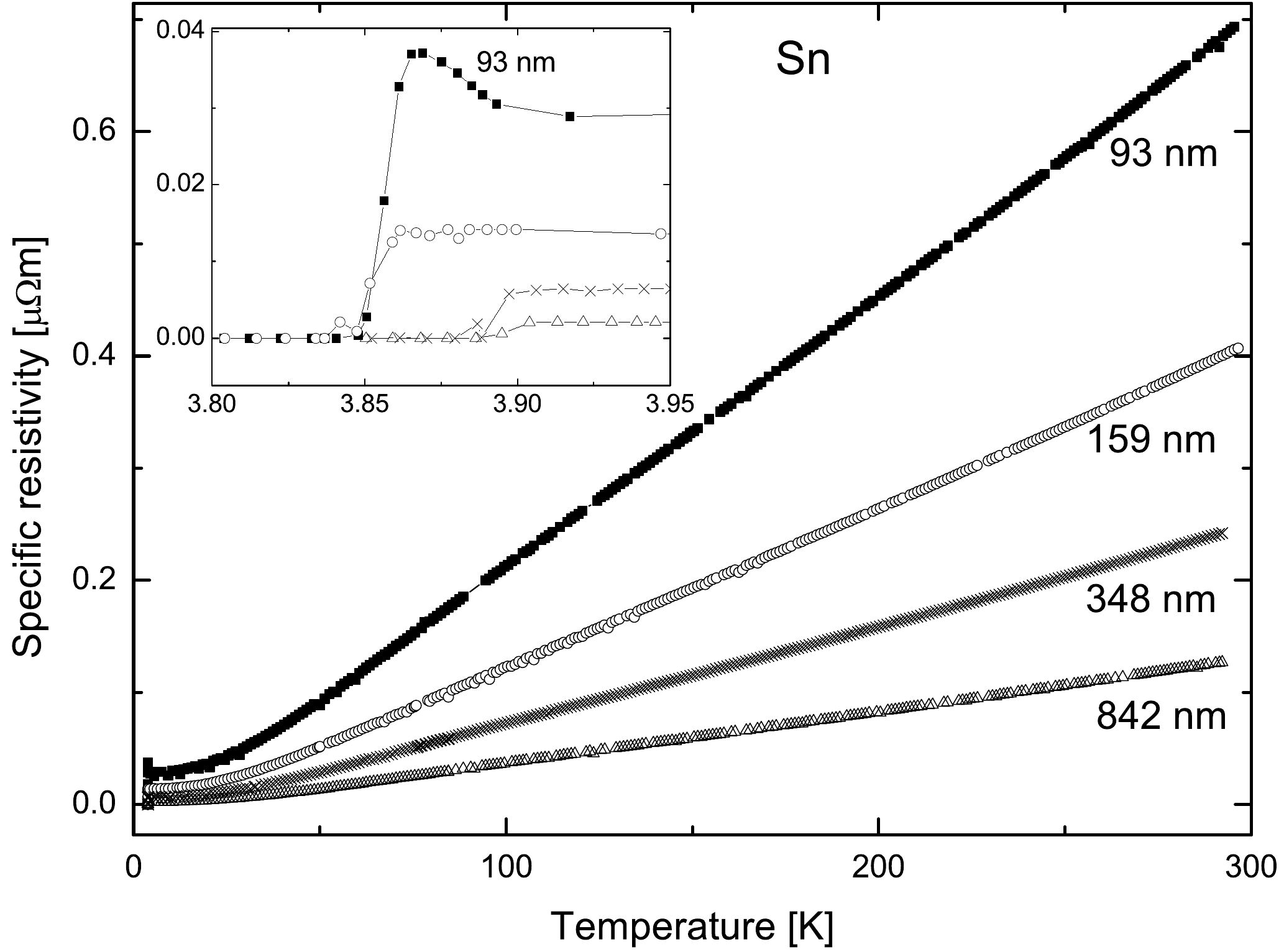}
\caption{Temperature-dependent DC resistivity for Sn films with different thickness. The inset shows the superconducting transition around $T_{c,\mathrm{Sn}}$.}
\label{fig:resistivity}
\end{figure}

The Sn films were thermally evaporated onto sapphire substrates (which were kept at room temperature) and are characterized using AFM and DC measurements as shown in Fig.\ \ref{fig:afm} and \ref{fig:resistivity}. The AFM images in Fig.\ \ref{fig:afm} demonstrate the expected behavior of Sn for different thickness starting from disconnected islands at $\unit[68]{nm}$ to a labyrinth structure at $\unit[93]{nm}$ and a connected surface consisting of a superposition of islands at $\unit[348]{nm}$. The film thickness was determined using a profilometer. 

DC measurements were performed in a $^4\mathrm{He}$ bath cryostat using the four probe van der Pauw technique \cite{vanderPauw_1952}. It allows the calculation of the DC resistivity of the sample independently of the actual sample shape. The DC measurements of samples with thickness $\unit[68]{nm}$ and lower resulted in not measurable high resistance which is in good agreement with the isolated structure seen in Fig.\ \ref{fig:afm}(a). Thicker samples, $d\geq\unit[93]{nm}$, show metallic behavior with enhanced conductivity for increasing film thickness and with resistivity that is basically linear in temperature between $\unit[40]{K}$ and $\unit[300]{K}$, as shown in Fig.\ \ref{fig:resistivity}.
For these films, the phase transition into the superconducting state can easily be detected in DC transport, as shown in the inset of Fig.\ \ref{fig:resistivity}, at a critical temperature between $T_c \approx \unit[3.85]{K}$ and $T_c \approx \unit[3.90]{K}$. (The origin of the upturn above $T_c$ for the 93~nm sample is not clear to us but apparently intrinsic whereas the slight differences in $T_c$ between samples might be due to uncertainty in temperature measurement.) 
These DC measurements suggest an insulating and a metallic regime for samples below and above percolation, respectively.

\section{Methods}
\begin{figure*}[tbph]
\centering
\begin{minipage}{0.4\textwidth}
\begin{center}
\includegraphics[width=1.2\linewidth]{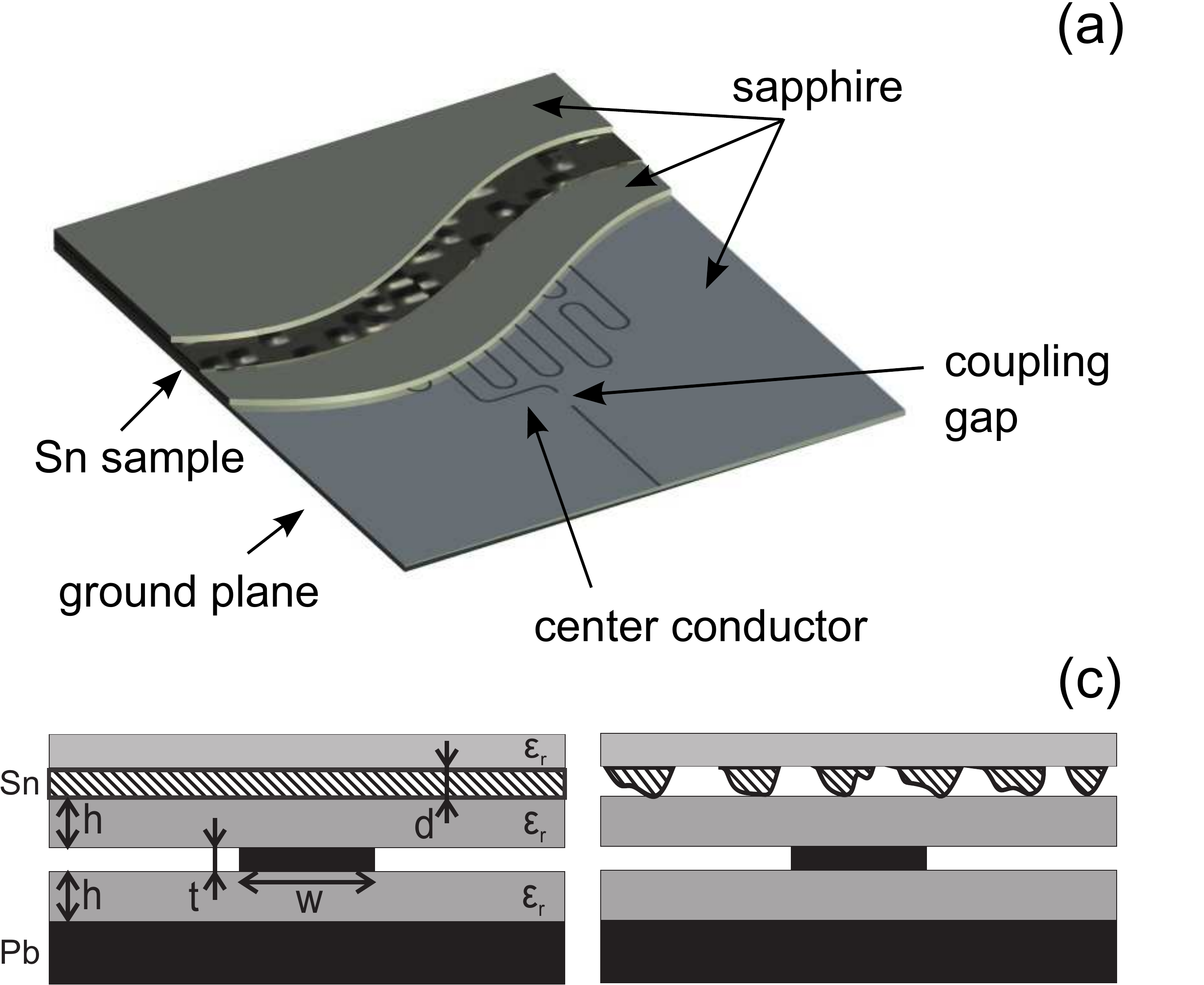}
\end{center}
\end{minipage}
\hspace{40pt}
\begin{minipage}{0.5\textwidth}
\begin{center}
\includegraphics[width=1\linewidth]{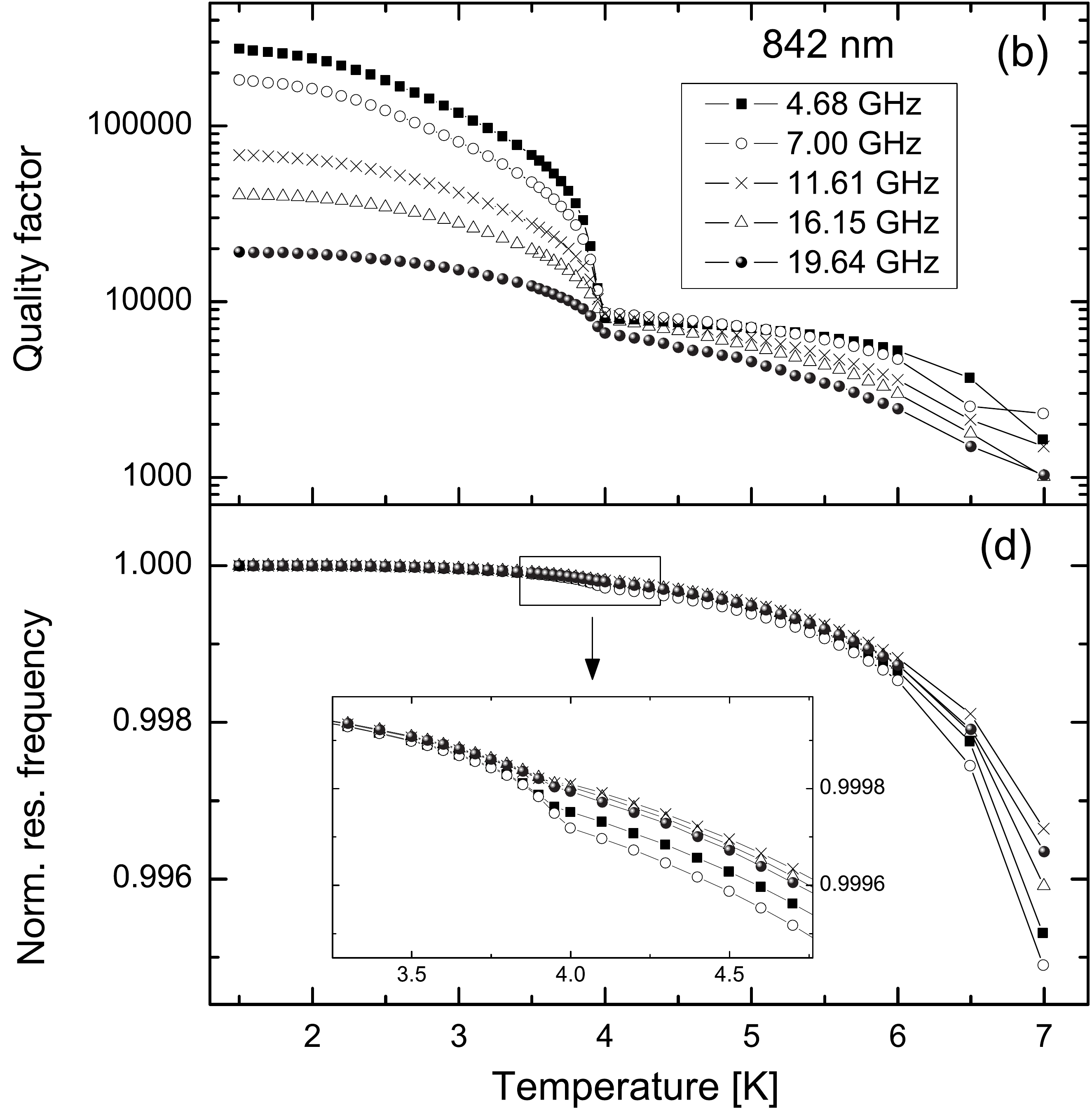}
\end{center}
\end{minipage}
\caption{Schematic 3D view (a) and cross-section (c) of the stripline resonator. Temperature-dependent quality factor (b) and resonance frequency (d) for a $\unit[842]{nm}$ thick Sn film at frequencies between $\unit[4.5]{GHz}$ and $\unit[20]{GHz}$. For reasons of comparability the resonance frequencies are normalized to their maximum value.}
\label{fig:schematic_freqdependence}
\end{figure*}

The design of the stripline resonators is shown schematically in Fig.\ \ref{fig:schematic_freqdependence}(a,c). They are based on a planar geometry and consist of a flat strip acting as center conductor which is sandwiched between conductive ground planes below and above. 
These layers are separated by sapphire substrates as dielectrics (see Fig.\ \ref{fig:schematic_freqdependence}(a,c)). The center conductor acts as one-dimensional transmission line and by fabricating two gaps at distance $l$, which reflect most of the microwave signal, a resonator with fundamental frequency of $\unit[4.5]{GHz}$ is formed. The resonator can easily be operated at higher harmonics up to $\unit[20]{GHz}$ due to its one-dimensionality \cite{hafner_2014}. The resonator is deposited on a $\unit[10\mathrm{x}12]{mm^2}$ sapphire chip and has a center conductor width of $W=\unit[60]{\mu m}$ and height $t=\unit[1]{\mu m}$. The surrounding sapphire substrate has a thickness of $\unit[125]{\mu m}$ and an effective dielectric constant $\epsilon_\mathrm{eff}\approx9.9$~\protect\footnotemark. With a coupling gap of $d_\mathrm{gap}=\unit[80]{\mu m}$, the resonator is undercoupled \cite{goeppl_2008,hafner_2014}.

The damping of the microwave signal depends strongly on the properties of the sample replacing the upper ground plane of the resonator.
The transmitted signal is mostly damped by the sample if a superconducting center strip with comparably lower losses is used, e.g.\ Pb with critical temperature $T_{c,\mathrm{Pb}} \approx \unit[7.2]{K}$. Below $T_{c,\mathrm{Pb}}$ the microwave losses in the center conductor are much lower compared to the dissipation caused by the sample. If the material of interest is also a superconductor, this property is maintained if the sample has a considerably lower $T_c$ than the center strip, like in our case $T_{c,\mathrm{Sn}} \approx \unit[3.8]{K}<T_{c,\mathrm{Pb}}$.

The measured transmission of the stripline resonator shows maxima when the electromagnetic wave meets the resonance condition $\nu_k=kc/(2\sqrt{\epsilon}l)=k\nu_0$ with fundamental resonance frequency $\nu_0$, permittivity $\epsilon$ of the dielectric and integer $k$. The resonances vanish if the temperature exceeds $T_{c,\mathrm{Pb}}$, since the losses in the normal-conducting phase of Pb are too high (for weak coupling). The measurements were performed in a $^{4}\mathrm{He}$ bath cryostat reaching down to temperatures well below $\unit[2]{K}$ and the complex transmission parameter $S_{21}$ was recorded by a vector network analyzer (VNA). 
In order to correct the collected spectra, a reference spectrum at a temperature slightly above $T_c$ of Pb, i.e. $\unit[7.5]{K}$ was measured and served for calibration purposes. For this temperature resonances are not observable and therefore one can subtract the residual background from the collected spectra below $T_{c,\mathrm{Pb}}$ \cite{hafner_2014}. The corrected signal shows a Lorentzian form which allows the calculation of the quality factor $Q=\nu_k/\Delta \nu_k$ with resonance frequency $\nu_k$ and full width at half maximum $\Delta \nu_k$ of the $k$-th mode. 
Figure\ \ref{fig:schematic_freqdependence}(b,d) shows the temperature-dependent behavior of resonance frequency and quality factor and demonstrates the broad frequency range which can be covered using superconducting Pb stripline resonators. The first decrease in quality factor and resonance frequency, below 4~K, corresponds to increasing losses of the Sn sample and the second one, at higher temperatures, to the resonator going normal-conducting.
The evolution of resonance frequency and quality factor when approaching $T_{c,\mathrm{Sn}}$ resembles the general course that has been observed previously in numerous microwave studies of superconducting materials \cite{Huttema2006, dilorio_stripline_1988,hafner_2014,Truncik_natcommns,Langley_revsciinstrum,krupka_eanalysis}, and which can consistently be explained by losses that are increasing with temperature.
To discuss the observed phenomena in terms of the fundamental material properties of a superconductor, such as conductivity or penetration depth, further analysis steps to the raw microwave data of Q and $\nu_k$ are necessary, and have been performed in related studies before \cite{dilorio_stripline_1988,hafner_2014,Anlage_jsupercond,Salluzzo_physrevlett85,Truncik_natcommns,krupka_eanalysis}. Also for our stripline geometry, this is conceptually possible, but impeded in the present work on Sn films close to percolation due to two main reasons: firstly, the Pb resonator still is somewhat temperature-dependent in the interesting temperature range between 1.5~K and slightly above $T_{c,\mathrm{Sn}}$, which contributes a temperature-dependent background. Secondly, the different Sn samples cross different thickness regimes from much thicker to eventually thinner than the penetration and skin depths, which does not allow for a universal analytical data treatment. Instead numeric procedures could be applied as further analysis step, but in the present work we do not follow this approach as the relevant effects for our discussion are already clearly observable in the raw data.

\section{Results and Discussion}
\begin{figure*}[tbph]
\centering
\includegraphics[width=1\linewidth]{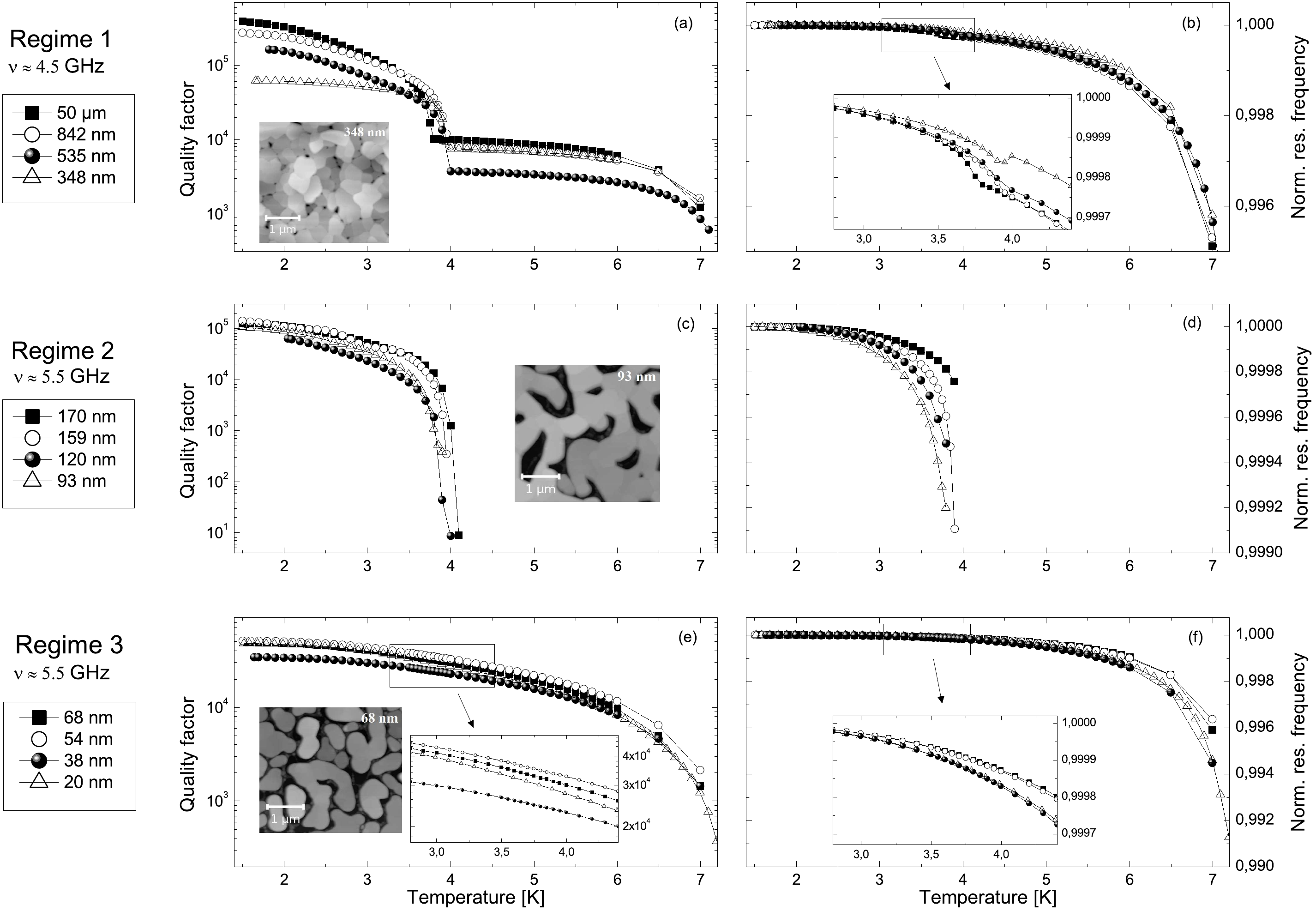}
\caption{In dependence of the film thickness there are three different regimes observable. a) Temperature-dependent quality factor $Q$ for samples $d\ge \unit[348]{nm}$. It declines with increasing temperatures caused by losses emerging in the resonator. Close to $T_{c,\mathrm{Sn}}\approx\unit[3.8]{K}$ the samples enter the normal-conducting phase resulting in a distinct kink in the course of $Q$; at temperatures up to $T_{c,\mathrm{Pb}}\approx\unit[7.2]{K}$ the quality factor decreases due to Pb approaching its metallic state. c) Quality factor for sample thicknesses $\unit[93]{nm}\le d\le\unit[170]{nm}$. The effects of the samples are still observable, however, in their normal-conducting state the quality factor drops rapidly due to further loss mechanisms. e) Quality factor for samples $d\le \unit[68]{nm}$. No effects of the samples on the properties of the resonator are observed. The decrease of quality factor is solely caused by the Pb resonator. b), d) and f) show the respective temperature-dependent resonance frequencies for each regime. Due to the penetration depth increasing with temperature the effective geometry of the resonator changes and so there is a decrease of the resonance frequencies. 
Samples with $d\ge \unit[348]{nm}$ additionally show a distinct kink at $T_{c,\mathrm{Sn}}\approx\unit[3.8]{K}$ (Inset). For samples in the second regime $\unit[93]{nm}\le d\le\unit[170]{nm}$ no resonances for $T>T_{c,\mathrm{Sn}}$ are observed, therefore the resonance frequency cannot be determined.}
\label{fig:stapelq}
\end{figure*}

Using the superconducting Pb  stripline resonators we studied several Sn samples, varying in thickness, for temperatures from $\unit[7.2]{K}$ (just blow $T_{c,\mathrm{Pb}}$) down to $\unit[1.5]{K}$. 
Figure\ \ref{fig:stapelq} shows quality factor and resonance frequency for several samples covering a wide range of thickness. All data were obtained at around $\unit[5]{GHz}$. 
There are three different thickness regimes in which the properties of the resonator ($Q$ as measure of losses and resonator frequency as a measure of microwave penetration) differ significantly from each other. 

Fig.\ \ref{fig:stapelq}(a) shows the general behavior of the quality factor $Q$ for thick samples, $d \ge \unit[348]{nm}$.
With increasing temperature there is a decrease of quality factor up to $T_{c,\mathrm{Sn}} \approx \unit[3.8]{K}$ originating from increasing absorption of the microwave signal, i.e. by thermally excited quasiparticles. For higher temperatures $Q$ is dominated by losses in the now metallic Sn sample, which in this range is basically temperature independent, and by losses in the Pb center conductor, where thermally excited quasiparticles lead to further decreasing $Q$ when approaching $T_{c,\mathrm{Pb}}$. In this first regime (closed, comparably thick films) resonances are observable both in the superconducting state of the Sn sample as well as in its normal-conducting state. The corresponding temperature-dependent resonance frequencies are depicted in Fig \ref{fig:stapelq}(b). For reasons of clarity and comparability, they are normalized to their maximum value. Increasing temperature leads to a decrease in resonance frequency with a distinct kink at $T_c$ of Sn (see inset).
The behavior of the resonators concerning quality factor and resonance frequency is explained by considering the penetration depth $\lambda(T)\propto 1/\sqrt{n(T)}$ which increases for higher temperatures since the density $n$ of Cooper pairs decreases. According to \cite{hafner_2014} the resonance frequency $\nu_k(T) = \nu_{k,T=0}  [1/(1+\lambda(T) \pi \mu_0/\Gamma)]^{1/2}$ of a stripline resonator is related to the penetration depth $\lambda(T)$, a structural constant $\Gamma$ depending on the resonator geometry, and the magnetic permeability $\mu_0$. As reference for bulk Sn also a $d=\unit[50]{\mu m}$ thick Sn foil was measured exhibiting behavior similar to samples ranging in thickness down to $\unit[348]{nm}$. 
This indicates for samples in this first regime bulk-like charge transport through a macroscopically connected film, which is in good agreement with the AFM pictures taken at $d=\unit[348]{nm}$ (Fig.\ \ref{fig:afm}(c)).

For samples $d \le \unit[68]{nm}$ the resonator exhibits no observable influence of the Sn sample. In this third regime the thickness is below percolation and the quality factor as well as the resonance frequency decrease continuously as shown in Fig.\ \ref{fig:stapelq}(e) and \ref{fig:stapelq}(f). In particular we do not find any features in the temperature dependence around $T_{c,\mathrm{Sn}}$.
From the AFM images (Fig.\ \ref{fig:afm}(a)) we know that these films consist of isolated islands. Thus these small metallic regions are not electrically connected to ground potential, and so they cannot act as ground plane for the stripline geometry. Therefore, resonators with such films more resemble a microstrip geometry, which consists of a signal-carrying strip and just one ground plane, and which is well-documented for application as superconducting resonators as well \cite{Scalapino_microstrip_1996,S.Anlage_microstrip_1992,B.Langley_microstrip(ausSchefflDiss)_1991}.
These observations demonstrate that macroscopic, connected current paths are not only required for finite DC conductivity of Sn, but also for its microwave conductivity. 

For Sn films in the intermediate regime ($\unit[93]{nm}\le d\le\unit[170]{nm}$) just above percolation, the properties of the resonator differ significantly from samples of the first and third regimes, in particular between $T_{c,\mathrm{Sn}}$ and $T_{c,\mathrm{Pb}}$ (see Fig.\ \ref{fig:stapelq}(c) and \ref{fig:stapelq}(d)). 
For temperatures well below $T_{c,\mathrm{Sn}}$ the quality factor as well as the resonance frequency exhibit similar behavior to samples of the first regime (Fig.\ \ref{fig:stapelq}(a,b)).
However, close to $T_{c,\mathrm{Sn}}$ the resonance frequency shows a more pronounced decrease when approaching the critical temperature of Sn compared to samples of the first regime. The resonances in the measured transmission spectra diminish and the quality factor decreases dramatically indicating  that the loss mechanisms in the Sn sample dominate the resonator performance. For $T>T_{c,\mathrm{Sn}}$ the resonances vanish and therefore the quality factor drops below measurable values. This rather disordered metallic state of Sn is too lossy for resonator operation. In \cite{Pal_increaseResistivity_1971,Mayadas_grain_boundary_scattering_1970} the increased resistivity is explained by considering, additionally to the normal background scattering caused by dislocations and phonons, scattering of conduction electrons by grain boundaries.

\begin{figure}[tbph]
\centering
\includegraphics[width=\linewidth]{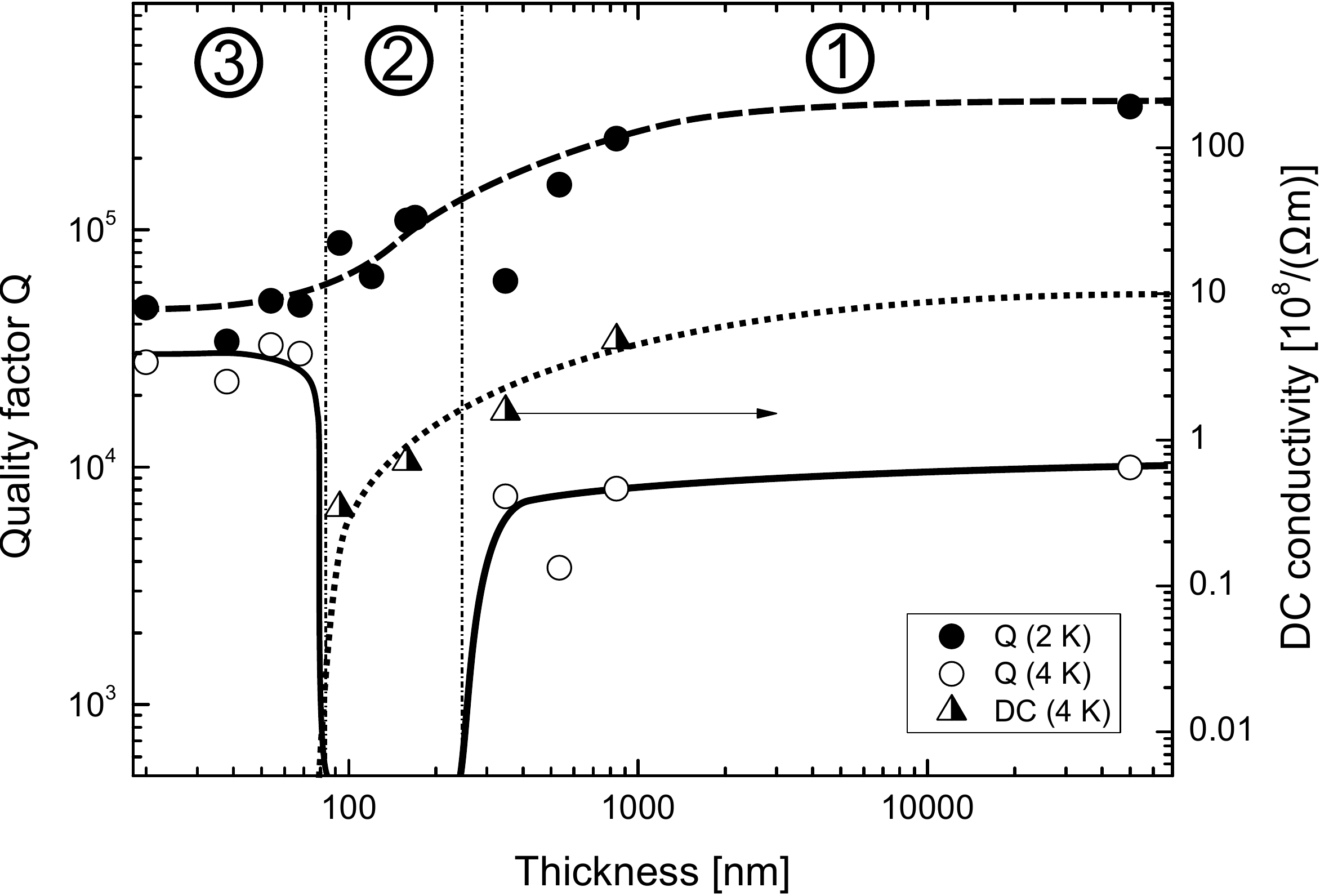}
\caption{Comparison of the quality factor $Q$ and the conductivity $\sigma_\mathrm{DC}$ obtained by microwave and DC measurements, respectively. In the normal-conducting phase of Sn (at $\unit[4]{K}$) the microwave measurements exhibit three different regimes (denoted with 1, 2 and 3) whereas the DC measurements suggest two regimes only, i.e. one insulating and one conductive regime. Lines are guides to the eye}
\label{fig:mv_DC}
\end{figure}

The overall picture concerning the electrodynamic response of the Sn films that we obtain from our experiments is summarized in Fig.\ \ref{fig:mv_DC}: for the normal-state DC conductivity, $\sigma_\mathrm{DC} = 1/\rho_\mathrm{DC}$, simple behavior as function of thickness is observed as expected, namely at thicknesses below percolation, $d<d_c \approx\unit[80]{nm}$, no macroscopic current transport is observed. For $d>d_c$, on the other hand, a superconducting state with $\rho_\mathrm{DC} =0$ is found below $T_c$ whereas $\rho_\mathrm{DC}$ above $T_c$ strongly depends on thickness. In particular, the strong increase of normal-state conductivity $\sigma_\mathrm{DC}$ above $d_c$, shown as triangles in Fig.\ \ref{fig:mv_DC}, can be easily explained by electronic scattering at the surfaces and grain boundaries of the Sn films that are clearly visible in the AFM images and that becomes less important as the film thickness is increased. In other words, for DC transport there are only two thickness regimes.

In the microwave response, the situation is more complex, as an additional regime emerges for thicknesses right above percolation. For film thickness below percolation, regime~3, the observed resonator response with high quality factors and no visible signature of $T_{c,\mathrm{Sn}}$ indicates that these Sn films do not affect resonator performance much, i.e.\ they do not influence the microwaves that are present in the resonator. Thus the Sn films behave like an insulator, which on a microscopic scale would be surprising, as the AFM images indicate micron-sized Sn islands, and the presence of a metallic layer on the substrate can be seen with bare eye (i.e.\ on the length scale of visible light). However, the probing length scales for DC is the sample size, well beyond the island dimensions, and therefore no current path for DC transport exists, resulting in the non-conducting state found in DC measurements. The probing length scale for microwaves, which in the simplest case can be approximated by the wavelength or the device/resonator dimensions, again are much bigger than the island size and therefore the insulating sample contribution to the resonator performance in this thickness regime~3 can easily be explained.

For thicknesses right above percolation, regime~2 ($d_c < d < \unit[200]{nm}$), the continuously connected Sn regions, which allow macroscopic supercurrent transport, clearly influence the microwave resonator, as drastic changes at $T_{c,\mathrm{Sn}}$ are observed. Below $T_{c,\mathrm{Sn}}$, the high $Q$ indicates that Sn acts as superconducting, low-loss film for the microwaves. The role of the different length scales here is evident from the resonator frequency in Fig.\ \ref{fig:stapelq}(d): for temperatures above 3~K, approaching $T_{c,\mathrm{Sn}}$, the frequencies decrease in a pronounced fashion, much more than for the thicker films (inset of Fig.\ \ref{fig:stapelq}(b)). For regime~2, the film thickness is of the order of the penetration depth. Therefore, part of the microwave signal can leak through the film, into the dielectric above. Thus the mode volume is substantially bigger compared to films that are much thicker (regime~3, inset of Fig.\ \ref{fig:stapelq}(b), with no leakage into the backing dielectric), and therefore the resonator frequency is reduced. This effect becomes more pronounced as one approaches $T_{c,\mathrm{Sn}}$ and the penetration depth of the Sn increases further. As expected (Fig.\ \ref{fig:stapelq}(d)), this behavior becomes less pronounced as $d$ increases and substantially surpasses the penetration depth for all temperatures except for very close to $T_{c,\mathrm{Sn}}$.
For temperatures above $T_{c,\mathrm{Sn}}$, samples in regime~2 are metallic in DC transport, but electronic scattering at the surfaces and grain boundaries is so strong that the concomitant microwave losses suppress resonator operation completely. One might expect a relation between the mean free path $l_\mathrm{mfp}$ of the electrons and the thickness of the film causing the difference between regime~2 and regime~1. But with a mean free path of $l_\mathrm{mfp} \approx \unit[2.1]{nm}$ \cite{mfp} and film thickness $d>\unit[93]{nm}$ this transition does not seem to be the case.

At even larger thicknesses, in regime~1, and below $T_{c,\mathrm{Sn}}$, the films are much thicker than the penetration depth $\lambda=\unit[34]{nm}$ for $T=0$ and bulk \cite{kittel} and one now finds the behavior that one generally expects for a superconducting resonator: low losses and high $Q$, and a temperature-dependent shift of the resonator frequency (Fig.\ \ref{fig:stapelq}(b)) that can be explained by the penetration of the microwave: as one approaches $T_{c,\mathrm{Sn}}$ from below and finally passes it, the penetration depth in the Sn increases and then, at $T_\mathrm{c,Sn}$, crosses over to the metallic skin depth $\delta(\omega=\unit[10]{GHz},T=\unit[4]{K}, d=\unit[348]{nm}) \approx \unit[1.0]{\mu m}$, which is basically constant in the temperature range up to $T_{c,\mathrm{Pb}}$. Since the metallic skin depth strongly depends on the microwave frequency whereas the superconducting penetration does not (except for very close to $T_c$), one expects that the temperature-dependent behavior slightly depends on frequency. However, with our present resolution we did not observe such an effect.
The temperature evolution between $T_{c,\mathrm{Sn}}$ and $T_{c,\mathrm{Pb}}$ is explained by the temperature-dependent penetration depth of the Pb. Even in this regime, where the Sn is a normal metal, the microwave losses in the Sn are low enough to allow resonator operation with still substantial $Q$ of order $10^4$.

\section{Summary}

In this article we presented a study of thermally evaporated Sn films below and above percolation. Superconducting Pb stripline resonators were used to probe the microwave response of the samples resulting in three different regimes depending on the film thickness. This behavior was not expected based solely on data of DC measurements which exhibit only an insulating and a metallic regime. 
Microwave measurements exhibit an intermediate regime in which the normal-conducting state of Sn is too lossy for resonator operation whereas the superconducting state only has low microwave losses. This peculiar regime is caused by the interplay of film thickness and enhanced scattering.

\section*{Acknowledgments}

We thank B. Gompf, T. Picot, and U. Pracht for helpful discussions. Financial support by DFG and by Carl-Zeiss-Stiftung is thankfully acknowledged.\\
\newline
$\mathrm{\ddagger}$ The dielectric constant of sapphire ranges between $\epsilon_\mathrm{1\perp}\approx9.27$ and $\epsilon_\mathrm{1\parallel}\approx11.34$ depending on the orientation of the optical c-axis \cite{krupka_sapphire,krupka_sapphire_2}. The sapphire crystal used in this work was cut in r-plane and the center conductor of our resonators has a meander shaped pattern which results in a preferred direction (majority of center conductor oriented in one direction) leading to an effective dielectric constant which we assume to $\epsilon_\mathrm{eff}\approx9.9$. The change in dielectric constant within the measured temperature and frequency range is expected to be negligible \cite{krupka_sapphire,konaka_sapphire_1991,krupka_sapphire_2}.


\section*{References}

\begin{thebibliography}{unsrt}
\bibitem{dressel_buch} Dressel M and Gr{\"u}ner G 2002 \textit{Electrodynamics of Solids} Cambridge University Press

\bibitem{Schwartz2000}Schwartz A, Scheffler M and Anlage S M 2000 
Determination of the magnetization scaling exponent for single-crystal La$_{0.8}$Sr$_{0.2}$MnO$_3$ by broadband microwave surface impedance measurements 
\textit{Phys. Rev. B} \textbf{61} R870

\bibitem{Lunkenheimer2000}Lunkenheimer P, Schneider U, Brand R and Loidl A 2000 
Glassy Dynamics
\textit{Contemp. Phys.} \textbf{41} 15 

\bibitem{Bovtun2001}Bovtun V, Petzelt J, Porokhonskyy V, Kamba S, Yakimenko Y 2001
Structure of the dielectric spectrum of relaxor ferroelectrics
\textit{J. Eur. Ceram. Soc.} \textbf{21} 1307

\bibitem{scheffler_slow_drude_relaxation_2005} Scheffler M, Dressel M, Jourdan M and Adrian H 2005 Extremely slow Drude relaxation of correlated electrons \textit{Nature} \textbf{438} 1135

\bibitem{Krupka2006}Krupka J 2006 
Frequency domain complex permittivity measurements at microwave frequencies 
\textit{Meas. Sci. Technol.} \textbf{17} R55

\bibitem{dressel_scheffler_ann_phys_2006} Dressel M and Scheffler M 2006 Verifying the Drude response \textit{Ann. Phys.} \textbf{15} 535


\bibitem{Bonn1992}Bonn D A, Dosanjh P, Liang R and Hardy W N 1992 
Evidence for rapid suppression of quasiparticle scattering below Tc in YBa2Cu3O7−δ
\textit{Phys. Rev. Lett.} \textbf{68} 2390

\bibitem{Kokales2000}Kokales J D, Fournier P, Mercaldo L V, Talanov V V, Greene R L and Steven M. Anlage 2000
Microwave Electrodynamics of Electron-Doped Cuprate Superconductors 
\textit{Phys. Rev. Lett.} \textbf{85} 3696

\bibitem{Steinberg2008}Steinberg K, Scheffler M and Dressel M 2008
Quasiparticle response of superconducting aluminum to electromagnetic radiation
\textit{Phys. Rev. B} \textbf{77} 214517

\bibitem{Hashimoto2009}Hashimoto K et al. 2009 
Microwave Penetration Depth and Quasiparticle Conductivity of PrFeAsO$_{1−y}$ Single Crystals: Evidence for a Full-Gap Superconductor
\textit{Phys. Rev. Lett.} \textbf{102} 017002

\bibitem{Pompeo2010}Pompeo N, Silva E, Sarti S, Attanasio C, Cirillo C 2010
New aspects of microwave properties of Nb in the mixed state
\textit{Physica C} \textbf{470} 901

\bibitem{Klein1993}Klein O, Donovan S, Dressel M and Gr\"uner G 1993
Microwave Cavity Perturbation Technique: Part I: Principles
\textit{Int. J. Infrared Millim. Waves} \textit{14} 2423

\bibitem{Booth1994}Booth J C, Wu D H, Anlage S M 1994
A broadband method for the measurement of the surface impedance of thin films at microwave frequencies
\textit{Rev. Sci. Instrum.} \textbf{65} 2082

\bibitem{bolometric_approach_Turner_Broun_2004} Turner P J, Broun D M, Kamal S, Hayden M E, Bobowski J S, Harris R, Morgan D C, Preston J S, Bonn D A and Hardy W N 2004 Bolometric technique for high-resolution broadband microwave spectroscopy of ultra-low-loss samples \textit{Rev. Sci. Instrum.} \textbf{75} 124

\bibitem{Scheffler2005a}Scheffler M, Dressel M 2005 
Broadband microwave spectroscopy in Corbino geometry for temperatures down to 1.7 K
\textit{Rev. Sci. Instrum.} \textbf{76} 074702

\bibitem{Huttema2006}Huttema W A, Morgan B, Turner P J, Hardy W N, Zhou X, Bonn D A, Liang R, Broun D M 2006
Apparatus for high-resolution microwave spectroscopy in strong magnetic fields
\textit{Rev. Sci. Instrum.} \textbf{77} 023901


\bibitem{scheffler_broadband_planar/stripline_resonators_corbino_2012} Scheffler M, Schlegel K, Clauss C, Hafner D, Fella C, Dressel M, Jourdan M, Sichelschmidt J, Krellner C, Geibel C and Steglich F 2013 Microwave spectroscopy on heavy-fermion systems: Probing the dynamics of charges and magnetic moments \textit{Phys. Status Solidi B} \textbf{250} 439

\bibitem {scheffler_corbino_stripline_2015} Scheffler M, et al. 2015 Broadband Corbino spectroscopy and stripline resonators to study the microwave properties of superconductors \textit{Acta IMEKO} \textbf{4} 47-52

\bibitem{dilorio_stripline_1988} Dilorio M S, Anderson A C and Tsaur B.-Y. 1988 rf surface resistance of Y-Ba-Cu-O thin films \textit{Phys. Rev. B} \textbf{38} 7019-7022

\bibitem{Oates1990} Oates D E,  Anderson A C and  Mankiewich P M 1990 Measurement of the surface resistance of YBa$_2$Cu$_3$O$_{7-\mathrm{x}}$ thin films using stripline resonators \textit{J. Supercond.} \textbf{3} 251

\bibitem{Revenaz1994} Revenaz S,  Oates D E,  Labbé-Lavigne D,  Dresselhaus G and Dresselhaus M S 1994 Frequency dependence of the surface impedance of YBa$_2$Cu$_3$O$_{7-\delta}$ thin films in a dc magnetic field: Investigation of vortex dynamics \textit{Phys. Rev. B} \textbf{50} 1178 

\bibitem{Belk1996} Belk N,  Oates D E, Feld D A, Dresselhaus G and  Dresselhaus M S 1996 Frequency and temperature dependence of the microwave surface impedance of YBa$_2$Cu$_3$O$_{7-\delta}$ thin films in a dc magnetic field: Investigation of vortex dynamics \textit{Phys. Rev. B} \textbf{53} 3459

\bibitem{Oates2010} Oates D E, Agassi Y D and Moeckly B H 2010  Microwave measurements of MgB$_2$: implications for applications and order-parameter symmetry \textit{Supercond. Sci. Technol.} \textbf{23} 3

\bibitem{Scheffler2012}Scheffler M, Fella C and Dressel M 2012
Stripline resonators for cryogenic microwave spectroscopy on metals and superconductors
\textit{J. Phys. Conf. Ser.} \textbf{400} 052031

\bibitem{hafner_2014} Hafner D, Dressel M and Scheffler M 2014 Surface-resistance measurements using superconducting stripline resonators \textit{Rev. Sci. Instrum.} \textbf{85} 014702

\bibitem{markus_all_Nb_and_Ta_loaded_resonator_2014} Thiemann M, Bothner D, Koelle D, Kleiner R, Dressel M and Scheffler M 2014 Niobium stripline resonators for microwave studies on superconductors \textit{J. Phys.: Conf. Ser.} \textbf{568} 022043

\bibitem{HafnerJPS_2014} Hafner D, Dressel M, Stockert O, Grube K, v. L\"ohneysen H and Scheffler M 2014 Anomalous Microwave Surface Resistance of CeCu$_6$ \textit{JPS Conf. Proc.} \textbf{3} 012016

\bibitem{Parkkinen2015} Parkkinen K, Dressel M, Kliemt K, Krellner C, Geibel C, Steglich F and Scheffler M 2015 Signatures of Phase Transitions in the Microwave Response of YbRh$_2$Si$_2$ \textit{Physics Procedia} \textbf{75} 340

\bibitem{pfuner_opt_prop_2009} Pfuner F, Degiorgi L, Baturina T I, Vinokur V M and Baklanov M R 2009 Optical properties of TiN thin films close to the superconductor-insulator transition \textit{New J. Phys.} \textbf{11} 113017

\bibitem{liu_corbino_2011} Liu W, Kim M, Sambandamurthy G and Armitage N P 2011 Dynamical study of phase fluctuations and their critical slowing down in amorphous superconducting films \textit{Phys. Rev. B} \textbf{84} 024511

\bibitem{driessen_2012} Driessen E F C, Coumou P C J J, Tromp R R, de Visser P J and Klapwijk T M 2012 Strongly Disordered TiN and NbTiN s-Wave Superconductors Probed by Microwave Electrodynamics \textit{Phys. Rev. Lett.} \textbf{109} 107003

\bibitem{Pracht2012}Pracht U S, Scheffler M, Dressel M, Kalok D F, Strunk C and Baturina T I 2012
Direct observation of the superconducting gap in a thin film of titanium nitride using terahertz spectroscopy
\textit{Phys. Rev. B} \textbf{86} 184503

\bibitem{matsunaga_2013} Matsunaga R, Hamada Y I, Makise K, Uzawa Y, Terai H, Wang Z and Shimano R 2013 Higgs Amplitude Mode in BCS superconductors Nb$_{1-x}$Ti$_x$N Induced by Terahertz Pulse Excitation \textit{Phys. Rev. Lett.} \textbf{111} 057002

\bibitem{sherman_pracht_2015} Sherman D, Pracht U S, Gorshunov B, Poran S, Jesudasan J, Chand M, Raychaudhuri P, Swanson M, Trivedi N, Auerbach A, Scheffler M and Dressel M 2015 The Higgs mode in disordered superconductors close to a quantum phase transition \textit{Nature Physics} \textbf{11} 188

\bibitem{Kirkpatrick1973}Kirkpatrick S 1973
Percolation and Conduction
\textit{Rev. Mod. Phys.} \textbf{45}, 574

\bibitem{Deutscher1980}Deutscher G, Entin-Wohlman O, Fishman S and Shapira Y 1980
Percolation description of granular superconductors
\textit{Phys. Rev. B} \textbf{21} 5041

\bibitem{Golosovsky1992}Golosovsky M, Tsindlekht M and Davidov D 1992
Microwave propagation through superconductor-polymer composites
\textit{Phys. Rev. B} \textbf{46} 11439

\bibitem{Eisterer2003}Eisterer M, Zehetmayer M, Weber H W 2003
Current Percolation and Anisotropy in Polycrystalline MgB$_2$
\textit{Phys. Rev. Lett.} \textbf{90} 247002

\bibitem{Yamamoto2007}Yamamoto A, Shimoyama J-I, Kishio K and Matsushita T  2007 
Limiting factors of normal-state conductivity in superconducting MgB$2$: an application of mean-field theory for a site percolation problem
\textit{Supercond. Sci. Technol.} \textbf{20} 658

\bibitem{Cuppens2010}Cuppens J, Romero C P, Lievens P and Van Bael M J 2010
Superconductivity in Pb cluster assembled systems with different degrees of coagulation
\textit{Phys. Rev. B} \textbf{81} 064517


\bibitem{Clerc1990}Clerc J P, Giraud G, Laugier J M and Luck J M 1990
The electrical conductivity of binary disordered systems, percolation clusters, fractals and related models
\textit{Adv. Phys.} \textbf{39} 191

\bibitem{Dyre2000}Dyre J C and Schrøder T B 2000
Universality of ac conduction in disordered solids
\textit{Rev. Mod. Phys.} \textbf{72} 873

\bibitem{Hövel2010} H\"ovel M, Gompf B and Dressel M 2010 Dielectric properties of ultrathin metal films around the percolation threshold \textit{Phys. Rev. B} \textbf{81} 035402

\bibitem{andrew_critfield_1948} Andrew E R 1948 Critical Field Measurements on Superconducting Tin Foils \textit{Phys. Soc. London} \textbf{A62} 88-94

\bibitem{lock_1951} Lock J M 1951 Penetration of Magnetic Fields into Superconductors III. Measurements on Thin Films of Tin, Lead and Indium \textit{Proc. Roy. Soc.} \textbf{A208}, 391

\bibitem{blumberg1962} Blumberg R H and Seraphim D P 1962 Effect of Elastic Strain on the Superconducting Critical Temperature of Evaporated Tin Films \textit{J. Appl. Phys.} \textbf{33} 162

\bibitem{hall_1965} Hall P M 1965 Effect of Stress on the Superconducting Transition Temperature of Thin Films of Tin \textit{J. Appl. Phys.} \textbf{36} 8

\bibitem{increaseTc_1970} Man'kovski$\check{\mathrm{i}}$ K K, Komnik V V and Dmitrenko I M 1970 Critical temperature of superconducting transition of thin tin films \textit{Zh. Eksp. Teor. Fiz.} \textbf{59} 740-752

\bibitem{dolan1974} Dolan G J 1974 Critical Thicknesses for the Transition from Intermediate- to Mixed-State Behavior in Superconducting Thin Films of Pb, Sn, and In \textit{J. Low Temp. Phys.} \textbf{15} 112

\bibitem{eichele1981} Eichele R, Kern W and Huebener R P 1981 Superconductivity of Thin Films of Lead, Indium, and Tin Prepared in the Presence of Oxygen \textit{Appl. Phys.} \textbf{25} 95-104

\bibitem{vanderPauw_1952} Van der Pauw L J 1958 A Method of Measuring Specific Resistivity and Hall Effect of Discs of Arbitrary Shapes \textit{Philips Res. Repts.} \textbf{13} 1-9

\bibitem{goeppl_2008} G\"oppl M, Fragner A, Baur M, Bianchetti R, Filipp S, Fink J M, Leek P J, Puebla G, Steffen L and Wallraff A 2008 Coplanar waveguide resonators for circuit quantum electrodynamics \textit{J. Appl. Phys.} \textbf{104} 113904

\bibitem{krupka_sapphire} Krupka J, Geyer R G, Kuhn M and Hinken J H 1994 Dielectric Properties of Single Crystals of Al$_2$O$_3$, LaAlO$_3$, NdGaO$_3$, SrTiO$_3$, and MgO at Cryogenic Temperatures \textit{IEEE Trans. Microw. Theory Techn.} \textbf{42} 10

\bibitem{konaka_sapphire_1991} Konaka T, Sato M, Asano H and Kubo S 1991 \textit{J. Supercond.} \textbf{4} 4

\bibitem{krupka_sapphire_2} J, Derzakowski K, Tobar M, Hartnett J and Geyer R G 1999 Complex permittivity of some ultralow loss dielectric crystals at cryogenic temperatures \textit{Meas. Sci. Technol.} \textbf{10} 387-392

\bibitem{Anlage_jsupercond} Anlage S M, Langley B W, Snortland H J, Eom C B, Geballe T H and Beasley M R 1990 Magnetic Penetration Depth Measurements with the Microstrip Resonator Rechnique \textit{J. Supercond.} \textbf{3} 3

\bibitem{Salluzzo_physrevlett85} Salluzzo M, Palomba F, Pica G and Andreone A 2000 Role of Nd/Ba Disorder on the Penetration Depth of Nd$_{1+x}$Ba$_{2-x}$Cu$_3$O$_{7-\delta}$ Thin Films \textit{Phys. Rev. Lett} \textbf{85} 5

\bibitem{Truncik_natcommns} Truncik C J S, Huttema W A, Turner P J, \"Ozcan S, Murphy N C, Carri\`{e}re P R, Thewalt E, Morse K J, Koenig A J, Sarrao J K and Broun D M 2013 Nodal quasiparticle dynamics in the heavy fermion superconductor CeCoIn$_5$ revealed by precision microwave spectroscopy \textit{Nat. Commun.} \textbf{4} 2477

\bibitem{Langley_revsciinstrum} Langley B W, Anlage S M, Pease R F W and Beasley M R 1991 Magnetic penetration depth measurements of superconducting thin films by a microstrip resonator technique \textit{Rev.Sci. Instrum.} \textbf{62} 1801 

\bibitem{krupka_eanalysis} Krupka J, Wosik J, Jastrzebski C, Mazierska J and Zdrojek M 2013 Complex Conductivity of YBCO Films in Normal and Superconducting States Probed by Microwave Measurements \textit{IEEE Trans. Appl. Supercond.} \textbf{23} 1-11

\bibitem{Pal_increaseResistivity_1971} Pal A K, Sen P 1971 Resistivity and temperature coefficient of resistivity of tin films \textit{Journal of Materials Science} \textbf{12} 1472-1475


\bibitem{Scalapino_microstrip_1996} Dahm T and Scalapino D J 1996 Theory of intermodulation in a superconducting microstrip resonator \textit{J. Appl. Phys.} \textbf{82} 464

\bibitem{S.Anlage_microstrip_1992} Anlage S M and Wu D 1992 Magnetic Penetration Depth Measurements in Cuprate Superconductors \textit{J. Supercond.} \textbf{5} 395

\bibitem{B.Langley_microstrip(ausSchefflDiss)_1991} Langley B W, Anlage S M, Pease R F W and Beasley M 1991 Magnetic Penetration Depth Measurements of Superconducting Thin Films by the Microstrip Resonator Technique \textit{Rev. Sci. Instrum.} \textbf{62} 1801 

\bibitem{Mayadas_grain_boundary_scattering_1970} Mayadas A F and Shatzkes M 1970 Electrical-Resistivity Model for Polycrystalline Films: the Case of Arbitrary Reflection at External Surfaces \textit{Phys. Rev. B} \textbf{1} 1382

\bibitem{mfp} Tanuma S, Powell C J and Penn D R 2011 Calculations of electron inelastic mean free paths \textit{Surf. Interface Anal.} \textbf{43} 689

\bibitem{kittel} Kittel C 2005 \textit{Introduction to Solid State Physics} John Wiley and Sons, Inc




\end{thebibliography}
\end{document}